\documentclass[twoside]{article}

\usepackage{amsmath,amssymb}

\catcode`\@=11
\long\def\@makefntext#1{
\protect\noindent \hbox to 3.2pt {\hskip-.9pt  
$^{{\eightrm\@thefnmark}}$\hfil}#1\hfill}		

\def\thefootnote{\fnsymbol{footnote}}
\def\@makefnmark{\hbox to 0pt{$^{\@thefnmark}$\hss}}	
	
\def\ps@myheadings{\let\@mkboth\@gobbletwo
\def\@oddhead{\hbox{}
\rightmark\hfil\eightrm\thepage}   
\def\@oddfoot{}\def\@evenhead{\eightrm\thepage\hfil
\leftmark\hbox{}}\def\@evenfoot{}
\def\sectionmark##1{}\def\subsectionmark##1{}}

\oddsidemargin=\evensidemargin
\renewcommand{\thefootnote}{\fnsymbol{footnote}}

\newcounter{sectionc}\newcounter{subsectionc}\newcounter{subsubsectionc}
\renewcommand{\section}[1] {\vspace{12pt}\addtocounter{sectionc}{1} 
\setcounter{subsectionc}{0}\setcounter{subsubsectionc}{0}\noindent 
	{\tenbf\thesectionc. #1}\par\vspace{5pt}}
\renewcommand{\subsection}[1] {\vspace{12pt}\addtocounter{subsectionc}{1} 
	\setcounter{subsubsectionc}{0}\noindent 
	{\bf\thesectionc.\thesubsectionc. {\kern1pt \bfit #1}}\par\vspace{5pt}}
\renewcommand{\subsubsection}[1] {\vspace{12pt}\addtocounter{subsubsectionc}{1}
	\noindent{\tenrm\thesectionc.\thesubsectionc.\thesubsubsectionc.
	{\kern1pt \tenit #1}}\par\vspace{5pt}}
\newcommand{\nonumsection}[1] {\vspace{12pt}\noindent{\tenbf #1}
	\par\vspace{5pt}}

\newcounter{appendixc}
\newcounter{subappendixc}[appendixc]
\newcounter{subsubappendixc}[subappendixc]
\renewcommand{\thesubappendixc}{\Alph{appendixc}.\arabic{subappendixc}}
\renewcommand{\thesubsubappendixc}
	{\Alph{appendixc}.\arabic{subappendixc}.\arabic{subsubappendixc}}

\renewcommand{\appendix}[1] {\vspace{12pt}
        \refstepcounter{appendixc}
        \setcounter{figure}{0}
        \setcounter{table}{0}
        \setcounter{lemma}{0}
        \setcounter{theorem}{0}
        \setcounter{corollary}{0}
        \setcounter{definition}{0}
        \setcounter{equation}{0}
        \renewcommand{\thefigure}{\Alph{appendixc}.\arabic{figure}}
        \renewcommand{\thetable}{\Alph{appendixc}.\arabic{table}}
        \renewcommand{\theappendixc}{\Alph{appendixc}}
        \renewcommand{\thelemma}{\Alph{appendixc}.\arabic{lemma}}
        \renewcommand{\thetheorem}{\Alph{appendixc}.\arabic{theorem}}
        \renewcommand{\thedefinition}{\Alph{appendixc}.\arabic{definition}}
        \renewcommand{\thecorollary}{\Alph{appendixc}.\arabic{corollary}}
        \renewcommand{\theequation}{\Alph{appendixc}.\arabic{equation}}
        \noindent{\tenbf Appendix \theappendixc #1}\par\vspace{5pt}}
\newcommand{\subappendix}[1] {\vspace{12pt}
        \refstepcounter{subappendixc}
        \noindent{\bf Appendix \thesubappendixc. {\kern1pt \bfit #1}}
	\par\vspace{5pt}}
\newcommand{\subsubappendix}[1] {\vspace{12pt}
        \refstepcounter{subsubappendixc}
        \noindent{\rm Appendix \thesubsubappendixc. {\kern1pt \tenit #1}}
	\par\vspace{5pt}}

\newcommand{\textlineskip}{\baselineskip=13pt}
\newcommand{\smalllineskip}{\baselineskip=10pt}

\def\eightcirc{
\begin{picture}(0,0)
\put(4.4,1.8){\circle{6.5}}
\end{picture}}
\def\eightcopyright{\eightcirc\kern2.7pt\hbox{\eightrm c}} 

\newcommand{\copyrightheading}[1]
	{\vspace*{-2.5cm}\smalllineskip{\flushleft
	 }}

\def\abstracts#1#2#3{{
	\centering{\begin{minipage}{4.5in}\baselineskip=10pt\footnotesize
	\parindent=0pt #1\par 
	\parindent=15pt #2\par
	\parindent=15pt #3
	\end{minipage}}\par}} 

\newcommand{\bibit}{\nineit}

\renewenvironment{thebibliography}[1]
	{\frenchspacing
	 \ninerm\baselineskip=11pt
	 \begin{list}{\arabic{enumi}.}
	{\usecounter{enumi}\setlength{\parsep}{0pt}
	 \setlength{\leftmargin 12.7pt}{\rightmargin 0pt} 
	 \setlength{\itemsep}{0pt} \settowidth
	{\labelwidth}{#1.}\sloppy}}{\end{list}}

\newcounter{itemlistc}
\newcounter{romanlistc}
\newcounter{alphlistc}
\newcounter{arabiclistc}

\newcommand{\fcaption}[1]{
        \refstepcounter{figure}
        \setbox\@tempboxa = \hbox{\footnotesize Fig.~\thefigure. #1}
        \ifdim \wd\@tempboxa > 5in
           {\begin{center}
        \parbox{5in}{\footnotesize\smalllineskip Fig.~\thefigure. #1}
            \end{center}}
        \else
             {\begin{center}
             {\footnotesize Fig.~\thefigure. #1}
              \end{center}}
        \fi}

\newcommand{\tcaption}[1]{
        \refstepcounter{table}
        \setbox\@tempboxa = \hbox{\footnotesize Table~\thetable. #1}
        \ifdim \wd\@tempboxa > 5in
           {\begin{center}
        \parbox{5in}{\footnotesize\smalllineskip Table~\thetable. #1}
            \end{center}}
        \else
             {\begin{center}
             {\footnotesize Table~\thetable. #1}
              \end{center}}
        \fi}

\def\@citex[#1]#2{\if@filesw\immediate\write\@auxout
	{\string\citation{#2}}\fi
\def\@citea{}\@cite{\@for\@citeb:=#2\do
	{\@citea\def\@citea{,}\@ifundefined
	{b@\@citeb}{{\bf ?}\@warning
	{Citation `\@citeb' on page \thepage \space undefined}}
	{\csname b@\@citeb\endcsname}}}{#1}}

\newif\if@cghi
\def\cite{\@cghitrue\@ifnextchar [{\@tempswatrue
	\@citex}{\@tempswafalse\@citex[]}}
\def\citelow{\@cghifalse\@ifnextchar [{\@tempswatrue
	\@citex}{\@tempswafalse\@citex[]}}
\def\@cite#1#2{{$\null^{#1}$\if@tempswa\typeout
	{IJCGA warning: optional citation argument 
	ignored: `#2'} \fi}}

\def\pmb#1{\setbox0=\hbox{#1}
	\kern-.025em\copy0\kern-\wd0
	\kern.05em\copy0\kern-\wd0
	\kern-.025em\raise.0433em\box0}

\def\fnm#1{$^{\mbox{\scriptsize #1}}$}
\def\fnt#1#2{\footnotetext{\kern-.3em
	{$^{\mbox{\scriptsize #1}}$}{#2}}}

\def\fpage#1{\begingroup
\voffset=.3in
\thispagestyle{empty}\begin{table}[b]\centerline{\footnotesize #1}
	\end{table}\endgroup}

\def\runninghead#1#2{\pagestyle{myheadings}
\markboth{{\protect\footnotesize\it{\quad #1}}\hfill}
{\hfill{\protect\footnotesize\it{#2\quad}}}}
\font\tenrm=cmr10
\font\tenit=cmti10 
\font\tenbf=cmbx10
\font\bfit=cmbxti10 at 10pt
\font\ninerm=cmr9
\font\nineit=cmti9

\font\eightrm=cmr8

\def\qed{\hbox{${\vcenter{\vbox{			
   \hrule height 0.4pt\hbox{\vrule width 0.4pt height 6pt
   \kern5pt\vrule width 0.4pt}\hrule height 0.4pt}}}$}}

\renewcommand{\thefootnote}{\fnsymbol{footnote}}	

\begin{document}

\runninghead{ Pad\'e-Improved Estimate of Perturbative Contributions $\ldots$ } { Pad\'e-Improved Estimate of Perturbative Contributions to   Inclusive Semileptonic $\ldots$ }

\normalsize\textlineskip
\thispagestyle{empty}
\setcounter{page}{1}

\copyrightheading{}			

\vspace*{0.88truein}

\fpage{1}
\centerline{\bf 
Pad\'e-Improved Estimate of Perturbative Contributions to
}
\vspace*{0.035truein}
\centerline{\bf   Inclusive Semileptonic $b\to u$ Decays }
\vspace*{0.37truein}
\centerline{\footnotesize T.G. Steele}
\vspace*{0.015truein}
\centerline{\footnotesize\it Department of Physics \& Engineering Physics, University
of Saskatchewan, 116 Science Place}
\baselineskip=10pt
\centerline{\footnotesize\it Saskatoon, Saskatchewan,  S7N 5E2,
Canada}
\vspace*{10pt}
\centerline{\footnotesize M.R.\ Ahmady, F.A.\ Chishtie, V.\ Elias}
\vspace*{0.015truein}
\centerline{\footnotesize\it Department of Applied Mathematics, University of Western Ontario}
\baselineskip=10pt
\centerline{\footnotesize\it London, Ontario, N6A 5B7, Canada}

\vspace*{0.21truein}
\abstracts{
Pad\'e-approximant methods are used to estimate the three-loop perturbative contributions to the 
inclusive semileptonic $b \to u$ decay rate. These improved
estimates of the decay rate reduce the theoretical uncertainty in the 
extraction of  the 
CKM matrix element $|V_{ub}|$ from the measured inclusive
semileptonic branching ratio. 
}{}{}

\textlineskip			
\vspace*{12pt}			

\noindent
In this paper we briefly review the development of Pad\'e approximation techniques to QCD 
quantities satisfying a renormalization group equation,\cite{pade_corr} and the application of 
these techniques to the
estimate of three-loop contributions to the inclusive semileptonic 
$b \to u$ decay rate.\cite{pade_bu}

The QCD perturbative contributions to the inclusive semileptonic decay rate 
$\Gamma\left(b\to X_u\ell^-\bar \nu_\ell\right)$ are  known to two-loop order.\cite{ritbergen}    
The theoretical prediction of the decay rate is an interesting phenomenological quantity since it
  depends on the CKM matrix element $\left\vert V_{ub}\right\vert$.  
Moreover,   the  
two-loop
calculation is mainly sensitive to $m_b$  since the $b$  mass is much larger  
than final state particle masses ($m_u$, $m_\ell$), and  $m_c$ only enters the partial 
rate $b\to u\ell\bar \nu_\ell c\bar c$ or in virtual corrections.  
The $\overline{MS}$ scheme obviates  the poor convergence of the perturbative series in on-shell 
schemes, leading to the following two-loop result for the decay rate for 
five active flavours:\cite{ritbergen}  
\begin{gather}
\Gamma\left(b \rightarrow X_u \ell^-\overline{\nu}_\ell\right) =
Km_b^5(\mu) S\left[x(\mu),L(\mu)\right]
\label{basic_rate}\\
K \equiv G_F^2 |V_{ub}|^2/192\pi^3 ~,~
x(\mu)=\frac{\alpha(\mu)}{\pi}~ ,~ L(\mu) = \log(w)=
\log\left[\frac{m_b^2(\mu)}{\mu^2}\right]
\\
S[x,L]    =  1 + x\left(a_0 - a_1 L\right) + x^2\left(b_0 - b_1 L + b_2
L^2\right)\label{s_form} \\
a_0 = 4.25360~, \; a_1 = 5~,~
b_0 = 26.7848~, \; b_1 = 36.9902~, \; b_2 = 17.2917
\label{pert_coeffs}
\end{gather}
where $\mu$ represents the renormalization scale.

Three-loop corrections to (\ref{basic_rate}) are potentially significant, since at 
$\mu=m_b=m_b\left(m_b\right)\approx 4.2\,{\rm GeV}$ we find
\begin{equation}
\Gamma = K m_b^5\left[1+0.30+0.14\right]\quad .
\label{two_loop}
\end{equation}
\pagebreak
\textheight=7.8truein
\setcounter{footnote}{0}
\renewcommand{\thefootnote}{\alph{footnote}}
The general form of $S\left(x,L\right)$ which determines the three-loop order decay rate is
\begin{equation}
S[x,L]    =
1 + x\left(a_0 - a_1 L\right) + x^2\left(b_0 - b_1 L + b_2L^2\right)
+ x^3\left(c_0 -c_1 L + c_2L^2-c_3L^3\right)
\label{three_loop_form}
\end{equation}

The decay rate $\Gamma$ satisfies a renormalization-group (RG) equation which determines the three-loop 
coefficients
$\left\{c_1,~ c_2,~c_3\right\}$, but leaves the crucial $c_0$ coefficient undetermined.
\begin{gather}
0=\mu\frac{d\Gamma}{d\mu}=\left[\mu\frac{\partial}{\partial\mu}+
\gamma(x)m_b\frac{\partial}{\partial m_b}+\beta(x) \frac{\partial}{\partial x}
\right]\Gamma
\label{RG-eqn}\\
c_1 = 249.592~,\; c_2 = 178.755~,\; c_3 = 50.9144
\label{RG_coeffs}
\end{gather}
Pad\'e approximation methods can be used to estimate the $c_i$. A comparison of these estimates 
with  the RG-determined coefficients provides  a test of the estimation procedure we use, as well as an 
estimate of the uncertainty in the value of $c_0$  obtained via Pad\'e methods.

Pad\'e approximation methods are applied to a perturbation series of the form
\begin{equation}
S(x)=1+R_1 x+R_2 x^2+R_3x^3+\ldots
\label{pade_form}
\end{equation}
where $R_1$ and $R_2$ are known from a two-loop calculation.  
An asymptotic error formula\cite{samuel} for the Pad\'e predictions established in applications to QCD
 leads to the Pad\'e prediction of $R_3$.\cite{pade_bu}
\begin{equation}
R_3=\frac{2R_2^3}{R_1\left(R_1^2+R_2\right)}
\label{r3_pade}
\end{equation} 
A complication in  the case we are considering is that  $R_1$, $R_2$ and hence $R_3$ are 
implicitly functions of the quantity $w=m_b^2/\mu^2$
\begin{equation}
R_1(w)=a_0-a_1\log(w) 
\quad ,\quad
R_2(w)=b_0-b_1\log(w)+b_2\log^2(w) 
\label{known_R}
\end{equation}
The Pad\'e prediction of  the coefficients $c_i$ in (\ref{three_loop_form})  is obtained from a
least squares fit between the $w$ dependence of the Pad\'e prediction $R_3(w)$ and the 
perturbative form
\begin{equation}
c_0-c_1\log(w)+c_2\log^2(w)-c_3\log^3(w) \quad .
\end{equation}
Thus the Pad\'e prediction of the $c_i$ is obtained by minimizing the following expression, in which $R_3(w)$ is
estimated by substitution of (\ref{known_R}) into (\ref{r3_pade}): 
\begin{equation}
\chi^2\left(c_i\right)=\int\limits_0^1dw\left[
R_3(w)-
\left( c_0-c_1\log(w)+c_2\log^2(w)-c_3\log^3(w)  \right)
\right]^2\quad .
\label{chi2}
\end{equation}
The resulting Pad\'e estimates of  the three-loop coefficients $c_i$ are
\begin{equation}
 c_0 = 198.4~,~ c_1 = 260.6~,~ c_2 = 183.9~,~ c_3 = 48.64\quad .
\label{pade_predict}
\end{equation}
These Pad\'e estimates agree with the RG values (\ref{RG_coeffs}) for $\{c_1,c_2,c_3\}$ to better than 5\% accuracy, 
 suggesting  a  corresponding uncertainty for 
the Pad\'e-estimated value of $c_0$.  Similar 
or better accuracy in the Pad\'e estimates of RG-accessible 
coefficients has also been obtained in applications to QCD correlation functions and 
Higgs decay rates.\cite{pade_corr,higgs}

Using the eq.\ (\ref{pade_predict}) values of $c_i$, we find the three-loop 
Pad\'e estimate of the decay rate
exhibits  reduced 
renormalization-scale ($\mu$) dependence compared to the two-loop prediction.\cite{pade_bu}
The significance of the three-loop effects can be assessed by comparing 
the  two-loop decay rate 
(\ref{two_loop}) with the three-loop Pad\'e estimate of the 
decay rate at the renormalization scale $\mu=m_b$
\begin{equation}
\Gamma=K m_b^5\left(m_b\right)\left[ 1+0.30+0.14+0.08\right]\quad .
\end{equation}

However, the choice of renormalization scale $\mu=m_b$ is not necessarily  optimal.\cite{ritbergen}  
For an improved prediction
 we use the minimal sensitivity value of $\mu$ where $\Gamma$ is stable under $\mu$ variations. 
QCD inputs for obtaining this minimal-sensitivity prediction use the four-loop $\beta$ 
function\cite{beta} and
anomalous mass dimension\cite{mass} to evolve $\alpha$ and $m_b$ 
 numerically to the scale $\mu$ from the  values 
$\alpha\left(M_Z\right)$ and $m_b\left(m_b\right)=4.17\,{\rm GeV}$,\cite{chetyrkin} with matching conditions
through thresholds when necessary.\cite{match}
The minimal sensitivity value of the decay rate occurs  near the $\tau$ mass at
$\mu=1.775\,{\rm GeV}$, leading to the Pad\'e determination of the three-loop inclusive semileptonic
decay rate:
\begin{equation}
\frac{\Gamma}{K} = \left[ 5.1213 \, {\rm GeV}\right]^5 
\left[1 - 0.6455 + 0.2477 - 0.0143\right] 
          = 2071 \, {\rm GeV}^5 
\end{equation}
The theoretical uncertainties in this Pad\'e determination of the decay rate involve higher-order 
perturbative effects,  uncertainty in the Pad\'e determination of $c_0$, uncertainty in  
$\alpha\left(M_Z\right)$ and $m_b\left(m_b\right)$, and nonperturbative contributions, leading to 
an estimate of the decay rate\cite{pade_bu}
\begin{equation}
\frac{\Gamma\left(b\to X_u\ell^-\bar \nu_\ell\right)}{K} = \left( 2065 \pm 14\%\right)\,{\rm GeV^5}
\quad ,\quad K = G_F^2 |V_{ub}|^2/192\pi^3\quad ,
\end{equation}
from which  $\left\vert V_{ub}\right\vert$ can be extracted with 
7\% theoretical uncertainty.

\nonumsection{Acknowledgements}
\noindent
The authors gratefully acknowledge research funding from the 
Natural Science and Engineering Research Council of Canada (NSERC).

\nonumsection{References}

\end{document}